\documentclass[aps,floatfix,prd,twocolumn,nofootinbib]{revtex4}
\usepackage{graphicx} 
\usepackage{amsmath} 
\usepackage{amssymb}
\usepackage{subfigure}

\newcommand{\beq}{\begin{equation}} 
\newcommand{\eeq}{\end{equation}}
\newcommand{\dsl}{{\partial \hspace{-2.0mm}/}}
\newcommand{\Tr}{\mathrm{Tr}}

\newcommand{\MeV}{\;\mathrm{MeV}}

\newcommand{\Ref}[1]{Ref.~\cite{#1}}
\newcommand{\Refs}[1]{Refs.~\cite{#1}}
\newcommand{\Eq}[1]{Eq.~(\ref{#1})}
\newcommand{\Eqs}[1]{Eqs.~(\ref{#1})}
\newcommand{\Fig}[1]{Fig.~\ref{#1}}

\newcommand{\Lfchi}{\mathcal{L}^{\left(4\right)}_{\chi}}
\newcommand{\Lfd}{\mathcal{L}^{\left(4\right)}_{d}}
\newcommand{\Lschi}{\mathcal{L}^{\left(6\right)}_{\chi}}
\newcommand{\Lschid}{\mathcal{L}^{\left(6\right)}_{\chi d}}
\newcommand{\Kt}{K^\prime}
\newcommand{\cflbcs}{$\rm CFL_{BCS}$}
\newcommand{\cflbec}{$\rm CFL_{BEC}$}
\newcommand{\tscbcs}{$\rm 2SC_{BCS}$}
\newcommand{\tscbec}{$\rm 2SC_{BEC}$}

\def\dblone{\hbox{$1\hskip -1.2pt\vrule depth 0pt height 1.6ex width
  0.7pt \vrule depth 0pt height 0.3pt width 0.12em$}}

\begin{document}
    
\title{Role of two-flavor color superconductor pairing in a
  three-flavor Nambu--Jona-Lasinio model with axial anomaly} 
\author{H.~Basler}\affiliation{Institut f\"ur Kernphysik, Technische
Universit\"at Darmstadt, Germany} 

\author{M.~Buballa}\affiliation{Institut f\"ur Kernphysik, Technische
Universit\"at Darmstadt, Germany} 

\begin{abstract}
The phase diagram of strongly interacting matter is studied within a 
three-flavor Nambu--Jona-Lasinio model, 
which contains the coupling between chiral and diquark condensates
through the axial anomaly.
Our results show that it is essential to include the 2SC phase 
in the analysis. 
While this is expected for realistic strange quark masses,
we find that even for equal up, down, and strange bare quark masses,
2SC pairing can be favored due to spontaneous flavor-symmetry
breaking by the axial anomaly.
This can lead to a rich phase structure, including BCS- and BEC-like
2SC and CFL phases and new endpoints. 
On the other hand, the low-temperature critical endpoint, which was 
found earlier in the same model without 2SC pairing, 
is almost removed from the phase diagram and 
cannot be reached from the low-density chirally broken phase 
without crossing a preceding first-order phase boundary.
For physical quark masses no additional critical endpoint is found. 

\end{abstract}

\date{\today}

\maketitle

\section{Introduction}
\label{sec:introduction}

The phase diagram of strongly interacting matter is studied with great
effort, both experimentally and theoretically. From direct observations
we know that at low temperature and low chemical potential chiral
symmetry is spontaneously broken and hadrons are the relevant degrees
of freedom. There are strong indications from heavy-ion experiments
that this is changed at high temperatures where quarks and gluons become the 
relevant degrees of freedom and a so-called quark-gluon plasma (QGP) is
formed (see Ref.~\cite{BraunMunzinger:2007zz} for an overview). 
This is also confirmed by lattice simulations of quantum
chromodynamics (QCD) at nonvanishing 
temperature~\cite{Cheng:2009be,Aoki:2009sc}. These calculations
indicate that, for physical quark masses and vanishing chemical potential,
the hadronic phase and the QGP are connected by a smooth 
crossover~\cite{Aoki:2006we}.

At high densities and low temperatures, on the other hand, 
strongly interacting matter is expected to be a color superconductor,
where the quarks form Cooper pairs 
(for reviews, see~\cite{Rajagopal:2000wf, Alford:2001dt, 
Schafer:2003vz, Rischke:2003mt, Buballa:2003qv, Shovkovy:2004me, 
Alford:2007xm}). 
This can rigorously be shown for asymptotically high densities by 
applying weak-coupling techniques to 
QCD~\cite{Son:1998uk,Schafer:1999jg,Shovkovy:1999mr,Schafer:1999fe}.
Unfortunately, these methods fail at more ``moderate'' densities,
which are of phenomenological interest. Moreover, because of the 
``sign problem'', the regime of low temperature and nonvanishing chemical
potential cannot be studied within lattice QCD.
However, in model calculations one typically finds that the hadronic phase
is bordered by a first-order phase boundary in the low-temperature  
region~\cite{Asakawa:1989bq,Berges:1998rc}.
The combination of this result with the notion of the crossover at
zero chemical potential then leads to the standard picture of the phase 
diagram where the first-order phase transition ends at a critical end 
point (CEP). The latter has attracted considerable attention, as it is
potentially detectable in heavy-ion experiments~\cite{Stephanov:1998dy}.

It is possible, however, that the phase boundary of the hadronic phase
contains more than one CEP.  For instance, it was found within a
Nambu--Jona-Lasinio (NJL) model that imposing charge neutrality or the
inclusion of vector interactions weaken the first-order phase
transition and can lead to a second end point at low temperatures near
the chemical potential axis~\cite{Kitazawa:2002bc, Zhang:2008wx,
  Zhang:2009mk}. Beside these mechanisms, it has been shown in a
Ginzburg-Landau (GL) analysis~\cite{Hatsuda:2006ps, Yamamoto:2007ah,
  Baym:2008me} that the interaction between the chiral and diquark
condensates, induced by the axial anomaly, can also lead to such a
low-temperature endpoint.  This result is rather general, but it
strongly depends on the values of the GL coefficients,
which could not be determined within this framework. It is therefore
very interesting that it has recently been confirmed explicitly within
an NJL-model study~\cite{Abuki:2010jq}.  Thereby the authors have
included a $U_A(1)$-symmetry breaking interaction term which connects
chiral and diquark condensates.  This term plays a crucial role in the
GL analysis but is usually neglected in the NJL model.
In \Ref{Abuki:2010jq} it was found that it indeed leads to the
emergence of a second CEP if the coupling is sufficiently strong.

However, the calculations have been performed under the simplifying
assumption of three quark flavors with equal masses. In this case it
appears natural that color superconducting quark matter is realized in
the color-flavor locked (CFL) phase~\cite{Alford:1998mk}, where up,
down, and strange quarks are paired in a very symmetric way.  The
authors of Ref.~\cite{Abuki:2010jq} have therefore considered only one
common diquark condensate and one common chiral condensate in their
analysis. The same is true for the GL analysis
of~\Refs{Hatsuda:2006ps, Yamamoto:2007ah, Baym:2008me} in the case of
equal quark masses. Of course, the assumption of equal quark masses is
quite unrealistic, leading to the question whether the second CEP can
still be found in a scenario where the strange quark mass is
significantly larger than the masses of the up and down quarks. In
order to investigate this question, some generalizations of the model
are necessary: When the quark masses are different, the chiral
condensates will take different values for different flavors and also
the diquark condensates will depend on the flavor content of the
paired quarks.  In particular, at large differences between strange
and nonstrange quark masses, the CFL pairing will become
unfavored~\cite{Alford:1999pa,Buballa:2001gj,Oertel:2002pj} and a
two-flavor color superconductor (2SC)~\cite{Rapp:1997zu,Alford:1997zt}
will emerge, where only up and down quarks are paired.

In the following we will analyze the effect of these generalizations 
on the phase structure. Thereby it was our original motivation to
investigate the effect of realistic mass differences.
However, to our surprise, we find that even for equal bare quark masses
the 2SC phase can be favored due to the coupling of the 
chiral and diquark condensates by the axial anomaly. 
As a result we find the scenario of Ref.~\cite{Abuki:2010jq} 
never to be favored, not even for equal quark masses.

In the following, this will be discussed in detail. 
In Sec.~\ref{sec:model} we briefly introduce the model and the
parameters before presenting the numerical results in
Sec.~\ref{sec:results} and concluding in Sec.~\ref{sec:conclusions}.

\section{The model}
\label{sec:model}
\subsection{The Lagrangian}

We adopt the NJL-type Lagrangian of \Ref{Abuki:2010jq},
\begin{equation}
 \mathcal{L} = \bar{q} (i\dsl -\hat{m} + \gamma_0\mu ) q +
 \Lfchi + \Lfd + \Lschi + \Lschid\,,
 \label{eq:L}
\end{equation}
which is based on a frequently used Lagrangian
(e.g., \cite{Ruester:2005jc,Abuki:2005ms,Warringa:2006dk,Basler:2009vk}),
extended by the interaction term $\Lschid$.
It describes the dynamics of a quark field $q$
with three color (r,g,b) and three flavor
(u,d,s) degrees of freedom.
The current quark masses enter through
the diagonal mass matrix $\hat{m} = \text{diag}_f \left(m_u, m_d,
m_s\right)$ and $\mu$ is the quark chemical potential.
In the present analysis,
we do not impose electric or color charge neutrality constraints.
These constraints play an important role in compact stars and can  
lead to a rather complicated phase structure~\cite{Ruester:2005jc,
Abuki:2005ms,Warringa:2006dk,Basler:2009vk,Blaschke:2005uj}.
On the other hand, they are less important in heavy-ion collisions,
where the matter does not need to be locally neutral. 
As a first step, we therefore use a common chemical potential for all 
quarks, in order to keep the analysis simple.

The Lagrangian \Eq{eq:L} includes a four-point interaction in the
quark-antiquark channel,
\begin{equation}
 \Lfchi = G \sum_{a=0}^8 \left[\left(\bar{q}\tau_a
 q\right)^2 + \left(\bar{q} i \gamma_5 \tau_a q \right)^2 \right]\,,
\label{eq:Lqbarq}
\end{equation}
and a four-point interaction in the quark-quark channel, 
\begin{alignat}{1}
 \Lfd = H \sum_{i,j=1}^3 & \big[ \left(\bar{q} i \gamma_5
   t_i l_j C \bar{q}^T\right)\left( q^T C i \gamma_5 t_i l_j q
 \right)
 \nonumber\\
 +\; & \left(\bar{q} t_i l_j C \bar{q}^T\right) \left( q^T C
   t_i l_j q \right) \big]\,.
\label{Lqq}
\end{alignat}
Here $G$ and $H$ are dimensionful coupling constants, 
$C =i\gamma^2\gamma^0$ is the charge conjugation matrix,
and $\tau_a$ are the
Gell-Mann matrices in flavor space, extended by 
$\tau_0 = \sqrt{2/3} \dblone_f$.
The flavor and color structure of the quark-quark interaction
is generated by the antisymmetric matrices
\begin{eqnarray}
  t_1 &\equiv \tau_7\,,\qquad t_2 &\equiv -\tau_5\,,
  \qquad t_3 \equiv \tau_2\,,\qquad
\nonumber\\ 
  l_1 &\equiv \lambda_7\,,\qquad l_2 &\equiv -\lambda_5\,,\qquad l_3 
  \equiv \lambda_2\,, 
\label{eq:tl}
\end{eqnarray}
where $\lambda_a$ are the Gell-Mann matrices in color space.

The four-point interaction terms $\Lfchi$ and $\Lfd$ are symmetric
under $U\left(3\right)_R \times U\left(3\right)_L$ transformations in
flavor space. In order to break the $U\left(1\right)$ axial symmetry
we include the standard six-point interaction 
term~\cite{Kobayashi:1970ji, 't Hooft:1976fv}
\begin{equation}
 \Lschi = -K \;\left\{\mathrm{det}_f
   \left[\bar{q}\left(1+\gamma_5\right) q\right] + \mathrm{det}_f
   \left[\bar{q} \left(1-\gamma_5\right) q \right]\right\}\,,
 \label{eq:L6}
\end{equation}
which can be related to instanton effects.
This term connects three incoming right-handed fields with three
outgoing left-handed fields, and vice versa.
In principle, it can therefore couple to three 
quark-antiquark channels (e.g., $(\bar uu)(\bar dd)(\bar ss)$) 
as well as to one quark-antiquark channel together with one
diquark and one anti-diquark (e.g., $(ud)(\bar u\bar d)(\bar ss)$).
However, in Hartree approximation, which will be employed below,
$\Lschi$ only connects quark-antiquark condensates but no
diquark condensates. Since on the other hand the coupling between 
quark-antiquark and diquark condensates was found to be important in
the GL analysis of Refs.~\cite{Hatsuda:2006ps,
Yamamoto:2007ah, Baym:2008me}, the authors of \Ref{Abuki:2010jq}
have introduced a second six-point interaction. In our notation,
\Eq{eq:tl}, this term can be written as
\begin{align}
\label{eq:L6tilde}
 \Lschid = \frac{\Kt}{8} &
 \sum_{i,j,k=1}^3 \sum_\pm \biggl[ \left(\bar{q} t_i l_k
   (1\pm\gamma_5) C \bar{q}^T\right) \notag\\*
 &\left( q^T C (1\pm\gamma_5) t_j l_k q \right) \left(\bar{q}_i
   (1\pm\gamma_5) q_j\right)\biggr]\,,
\end{align}
where in the last factor the indices $i$ and $j$ refer to the flavor
components.  $\Lschid$ has the same symmetries as \Eq{eq:L6} but the
quark fields are ordered in such a way that in Hartree approximation
the diquark and quark-antiquark condensates are connected.  This
reordering of the quark fields can be realized via a Fierz
transformation,\footnote{Useful references in this
  context are~\Refs{Dmitrasinovic:2001nv, Nieves:2003in,
    Steiner:2005jm}.}  
of the instanton vertex, relating $\Lschi$ to $\Lschid$. 
In this way one can also relate the new coupling constant $\Kt$ to
$K$. However, following \Ref{Abuki:2010jq}, we will treat $\Kt$ as a 
free independent parameter.
 
\subsection{Mean-field approximation and thermodynamic potential}
We are working in mean-field approximation by introducing the
scalar diquark condensates
\begin{equation}
s_i = \langle q^T C \,\gamma_5\,t_i\,l_i\,q \rangle
\end{equation}
and the scalar antiquark-quark condensates
\begin{equation}
\phi_i = \langle \bar{q}_i q_i \rangle\,,\qquad i = 1,2,3\,.
\end{equation}
In Hartree approximation the six-point interaction $\Lschid$,
\Eq{eq:L6tilde}, then is simplified to
\begin{align}
 {\Lschid}_{MF} = \frac{\Kt}{4} \sum_{i=1}^3
 \Biggl[ -|s_i|^2\, \bar{q}_i q_i
 -s_i^{\star}\, q^T C \gamma_5\,t_i\, l_i\, q\, \phi_i \notag\\
 +\bar{q} \gamma_5 \,t_i \,l_i \,C \bar{q}^T
 \,s_i\, \phi_i
 +2\,|s_i|^2 \phi_i \Biggr]\,.
\end{align}
One can identify the first term as a contribution to the
effective quark masses, while the following two terms contribute to
the anomalous selfenergy of the propagator in the color
superconducting phase. The last term
does not depend on the fields and gives a constant contribution. 

Adding the other terms from \Eq{eq:L} and using Nambu-Gorkov bispinors 
$\Psi^T = 1/\sqrt{2} \left(q, \, C  \bar{q}^T \right)$
the full mean-field Lagrangian can be written as
\begin{equation}
\label{eq:LMF}
\mathcal{L}^{MF} = \bar{\Psi} S^{-1} \Psi - \mathcal{V},
\end{equation}
with the inverse dressed propagator
\begin{equation}
\label{eq:invProp}
S^{-1}(p) = \left(
\begin{array}{cc}
 p\hspace{-1.4mm}/ + \mu\gamma_0 - \hat{M} & \sum_{i=1}^3
\Delta_i \gamma_5 \, t_i \, l_i \\
-\sum_{i=1}^3 \Delta_i^{\star} \gamma_5\,t_i\,l_i &
p\hspace{-1.4mm}/ -\mu\gamma_0 - \hat{M}
\end{array}
\right)\,. 
\end{equation}
Here $\hat{M}$ is the diagonal mass matrix
of the constituent quark masses with the components
\begin{equation}
 M_i = m_i - 4G\phi_i + K|\epsilon_{ijk}|\phi_j\phi_k +
 \frac{\Kt}{4} |s_i|^2\,.
\label{eq:mass}
\end{equation}
The off-diagonal elements of \Eq{eq:invProp} include
\begin{equation}
\Delta_i = -2 \left(H - \frac{\Kt}{4}
 \phi_i \right) s_i\,.
\label{eq:delta}
\end{equation}
Finally $\mathcal{L}^{MF}$ includes a field-independent term
\begin{equation}
 \mathcal{V} = 2G \sum_{i=1}^3 \phi_i^2 -
 4K\phi_1\phi_2\phi_3 + \sum_{i=1}^3 \left(H - \frac{\Kt}{2}\phi_i
 \right) |s_i|^2\,. 
\label{eq:V}
\end{equation}
With these ingredients the thermodynamic potential in mean-field
approximation becomes
\begin{equation}
\Omega\left(T,\mu\right) = -T\sum_n \int \frac{d^3p}{\left( 2\pi
 \right)^3} \frac{1}{2} \Tr \ln \left(\frac{S^{-1}\left(
     i\omega_n,\vec{p}\right)}{T}\right) + \mathcal{V}\,,
\end{equation}
where the sum is over fermionic Matsubara frequencies. 
For equal diquark condensates, $s_1 = s_2 = s_3 \equiv s$, 
and equal chiral condensates, $\phi_1 = \phi_2 = \phi_3 \equiv \phi$,
it agrees with the thermodynamic potential of \Ref{Abuki:2010jq},
where only this limit was considered.

The self-consistent mean-field solutions are given by the 
stationary points of $\Omega$ with respect to the condensates,
i.e., by the solutions of the gap equations
\begin{equation}
\frac{\partial\Omega}{\partial\phi_i} =
\frac{\partial\Omega}{\partial s_i} = 0\,, \qquad i = 1,2,3\,.
\label{eq:gap}
\end{equation}
The phase diagram is then constructed by taking the solution
with the highest pressure (i.e., the lowest value of $\Omega$)
at each point in the $\mu$-$T$ plane.

\subsection{Parameters}
\label{ssec:parameters}

The model defined above contains eight parameters: the three bare quark masses
($m_u, m_d, m_s$), the two four-point coupling constants ($G,H$), the
two six-point coupling constants ($K,\Kt$), and a regularization parameter.
For the latter we take a sharp three-momentum cut-off $\Lambda$.

To fix these parameters, we basically follow the procedure of 
\Ref{Abuki:2010jq}. 
Starting point are the parameters of \Ref{Rehberg:1995kh},
\begin{align}
 m_u &= m_d = 5.5\MeV\,, \notag\\
 m_s &= 140.7\MeV\,, \notag\\
 \Lambda &= 602.3\MeV\,,\notag\\
 G &= 1.835/\Lambda^2\,,\notag\\
 K &= 12.36/\Lambda^5\,,
\label{paraphys}
\end{align}
which have been obtained by fitting meson masses and decay
constants in vacuum.
The value of the quark-quark coupling $H$, which cannot
be determined from vacuum meson properties, is taken
from~\Ref{Buballa:2003qv},
\begin{align}
 H & = 1.74/\Lambda^2\,.
\end{align}
Finally, as already mentioned, we treat the coupling constant $\Kt$
as a free parameter. We will often take $\Kt = 4.2\,K$,  
because with this value a second endpoint was found in~\Ref{Abuki:2010jq}.

The authors of \Ref{Abuki:2010jq} have only studied the case of equal quark
masses. To that end, they reduced the bare strange quark mass to the
value of the up and down quark masses
and then re-adjusted the antiquark-quark coupling $G$,
so that the dynamical up and down quark masses $M_u=M_d$ 
remain unchanged in vacuum (parameter set II of \Ref{Abuki:2010jq}): 
\beq
   m_s = m_u = m_d = 5.5\MeV\,, \quad G = 1.918/\Lambda^2\,.
\label{paraeq}
\eeq
In this way the vacuum values of the chiral condensates $\phi_u=\phi_d$ 
and of the pion mass and decay constant remain unchanged as well. 

Since we are interested in the effect of the strange quark mass on the
phase structure we generalize this procedure to the case of arbitrary
values of $m_s$.
Starting with equal quark masses, \Eq{paraeq},
we will increase the bare strange quark mass up to the more realistic
case, \Eq{paraphys}, 
always keeping the dynamical vacuum masses $M_u = M_d$ fixed
by adjusting the coupling constant $G$.

\section{Results}
\label{sec:results}
In this section we present our numerical results, which mainly
consist of a series of phase diagrams with different six-point
couplings $\Kt$ and different choices for the strange quark mass.
We begin with the case of
equal bare quark masses in Sec.~\ref{ssec:equalmasses},
before introducing larger values of $m_s$ in
Sec.~\ref{ssec:strangequarkmass}.

\subsection{Equal bare masses}
\label{ssec:equalmasses}

\begin{figure}
  \includegraphics[width=0.5\textwidth]{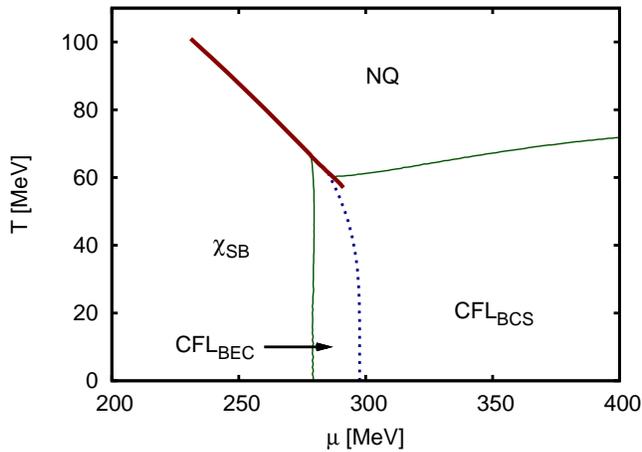}
  \caption{The phase diagram in the $\mu$-$T$ plane for $\Kt = 4.2\,K$
    and equal bare quark masses $m_u = m_d = m_s = 5.5\MeV$, only 
    allowing for one common diquark condensate $s$ 
    and one common chiral condensate $\phi$.
    Thick (red) solid lines denote first order phase transitions,
    thin (green) solid lines second order phase transitions.
    The dotted (blue) line indicates the BEC-BCS crossover line defined
    by $M(T,\mu) = \mu$, where $M\equiv M_u=M_d=M_s$.}
  \label{fig:phasediagramCFL}
\end{figure}

As a basis for our investigations, we consider equal quark masses,
$m_u = m_d = m_s = 5.5\MeV$ and allow for only one common chiral
condensate $\phi_1=\phi_2=\phi_3$ and one common diquark
$s_1 = s_2 = s_3$. For $\Kt = 4.2\,K$ we then reproduce the phase diagram
obtained in \Ref{Abuki:2010jq}, as shown in \Fig{fig:phasediagramCFL}.  
The normal
phases, where all diquark condensates vanish, are denoted by `$\chi$SB' in
regions with large chiral condensates and by `NQ' in regions with small
chiral condensates.
At low chemical potential and high temperature these two regions are 
connected by a crossover, while at higher chemical potential and lower
temperature they are separated by a first-order phase boundary. 
Without diquark condensates the latter would continue down to
the chemical-potential axis.  
However, when diquark condensates are included, the CFL phase is
favored in the lower right part of the phase diagram, replacing the
NQ phase in that regime.
The first-order chiral phase transition then does no longer go down 
to zero temperature but ends inside the CFL phase,\footnote{In this
article we use the term `chiral phase transition' whenever the transition 
is mainly characterized by a change of the chiral condensates $\phi_i$.
Note, however, that in the CFL phase chiral symmetry is always broken 
by the diquark condensates.}
while the latter is bordered to the $\chi$SB phase by a
second-order phase boundary.

As discussed in  \Ref{Abuki:2010jq}, this particular phase structure
is a consequence of the interaction term $\Lschid$.
Without this term, i.e., for $\Kt = 0$, the $\chi$SB-CFL phase
transition is first order and basically a continuation of the
first-order $\chi$SB-NQ phase transition 
(see \Fig{fig:phasediagrams5} below). In particular, the chiral
condensate drops considerably at the phase boundary and is
therefore small in the entire CFL phase.
This is different for sufficiently large values of $\Kt$, as in
\Fig{fig:phasediagramCFL}.
In this case the coupling between chiral and diquark 
condensates leads to the existence of strongly bound diquarks in the 
$\chi$SB phase, which eventually condense and then form a
Bose-Einstein condensate (BEC) in the lower-$\mu$ part of the 
CFL phase (see also \Ref{Kitazawa:2007zs}).
Hence, the $\chi$SB-CFL phase boundary is simply the condensation line
of the diquarks and therefore the transition is continuous.
In particular, the chiral condensate is still large at the phase boundary 
and only decreases considerably at somewhat higher values of $\mu$.
In a small regime near the boundary to the NQ phase 
this happens discontinuously and is just the continuation of the 
first-order phase transition between the $\chi$SB and the NQ phase.
At most temperatures, however, the decrease is continuous and closely 
related to a BEC-BCS crossover where the BEC-like CFL phase
(`\cflbec') gets converted into a BCS-like CFL phase (`\cflbcs').
The corresponding crossover line is indicated in the figure as well.
Following \Ref{Abuki:2010jq} we have defined it as the line where the 
in-medium constituent quark masses are equal to $\mu$. Note that with
this definition the crossover line does not exactly run into the
endpoint of the first-order phase boundary. 

\begin{figure}
  \includegraphics[width=0.5\textwidth]{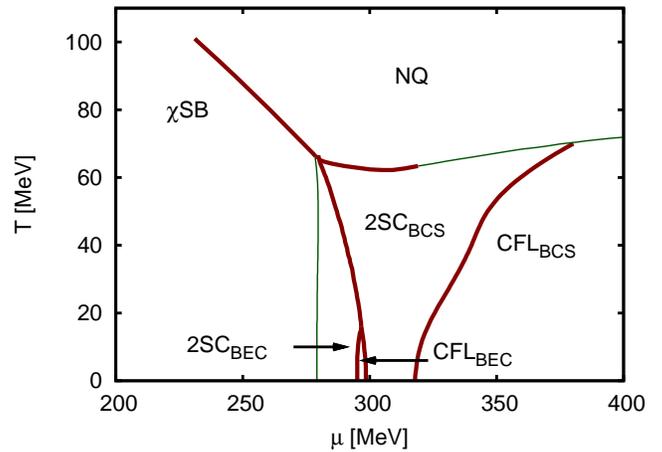}
  \caption{The same as \Fig{fig:phasediagramCFL} but with independent
    diquark condensates $s_1$, $s_2$, $s_3$,  
    and chiral condensates $\phi_1$, $\phi_2$, $\phi_3$.}
  \label{fig:phasediagram5.5}
\end{figure}

Next we allow the different diquark condensates $s_i$ and the chiral 
condensates $\phi_i$ to vary independently for $i=1,2,3$. This
opens the possibility for the formation of a 2SC phase ($s_1 = s_2 =
0,\,s_3\not=0$). Still keeping the bare quark masses equal, we obtain
the phase diagram shown in~\Fig{fig:phasediagram5.5}.  
Whereas the normal-conducting phases NQ and $\chi$SB 
occupy approximately the same regions of the phase diagram 
as in \Fig{fig:phasediagramCFL}, we find that a large part of the CFL 
phase is replaced by a 2SC phase. The first-order chiral phase transition,
which in \Fig{fig:phasediagramCFL} ends inside the CFL phase, now separates 
a BEC-like 2SC phase with large chiral condensates (`\tscbec')  
and a BCS-like 2SC phase with small chiral condensates (`\tscbcs').
However, unlike in \Fig{fig:phasediagramCFL},
the BEC-like and the BCS-like phases are always
separated by a first-order phase transition and
never connected by a crossover. 

Since $s_1$ and $s_2$ vanish in the 2SC phase whereas they are equal
to $s_3$ in the CFL phase, the two phases are necessarily separated by
first-order phase boundaries.
On the other hand, the phase transition from the CFL phase as well as
from the neighboring region of the 2SC phase to the NQ phase is of second
order.\footnote{For simplicity, we drop the subscripts `BCS' or `BEC'
when the distinction between BCS-like or BEC-like phases is not
relevant for the discussion.}
As an interesting consequence, the first-order 2SC-CFL 
phase-transition line ends exactly on the second-order phase boundary
to the NQ phase. Also note that the 2SC-NQ phase boundary, in the 
regime where it is second order, exactly agrees with the 
CFL-NQ phase boundary in \Fig{fig:phasediagramCFL}. 
This is due to the fact that on the second-order phase
boundary all diquark condensates vanish and therefore the 2SC phase
and the CFL phase have the same free energy. 
For similar reasons the second-order $\chi$SB-{\tscbec} phase boundary
agrees with the $\chi$SB-{\cflbec} phase boundary in 
\Fig{fig:phasediagramCFL}. 
Again, this is not surprising because both boundaries are related to
the Bose-Einstein condensation of diquarks, which 
are identical in the $\chi$SB phase and, hence, on the boundary.

At low temperatures ($T < 16\MeV$) we find a small area 
between the {\tscbec} phase and the {\tscbcs} phase, 
where the {\cflbec} phase is slightly preferred.
The origin of this ``CFL island'' will become more clear below. 

\begin{figure}
  \includegraphics[width=0.5\textwidth]{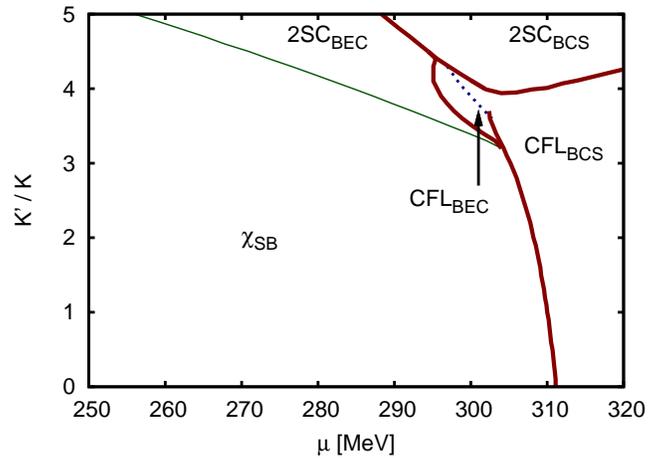}
  \caption{The phase diagram in the $\mu$-$\Kt$ plane at $T=0$ for
    equal quark masses, $m_u = m_d = m_s = 5.5\MeV$. 
    The meaning of the different line types is the same as in 
    \Fig{fig:phasediagramCFL}.}
  \label{fig:muKdi}
\end{figure}

In order to get a more complete picture, we now vary the
coupling constant $\Kt$ of the six-point interaction $\Lschid$.
In~\Fig{fig:muKdi} we show the phase diagram in the $\mu$-$\Kt$ plane 
for $T=0$. 
One immediately recognizes that for all choices of $\Kt$ there is
always a first-order phase transition at some value of $\mu$.
For low values of $\Kt$ only the $\chi$SB phase and the CFL phase
are present. The first interesting development occurs
when the interaction gets strong enough to have bound diquarks in the
$\chi$SB phase, which then condense at some value of $\mu$.
As a consequence, a {\tscbec} phase appears in the phase diagram.
At a slightly higher value of $\Kt$ the CFL phase
splits into a BEC-like and a BCS-like region.
These two regimes are separated by a first-order transition, which 
eventually turns into a crossover at higher values of $\Kt$.
The corresponding endpoint marks the critical value of $\Kt$ 
where in the phase diagram with equal condensates ($s_1 = s_2 = s_3$, 
$\phi_1 = \phi_2 = \phi_3$) the lower critical end point appears on the
$\mu$-axis. However, allowing for 2SC pairing, there still is a
first-order transition between the {\tscbec} and the {\cflbec} phase. 

Cutting \Fig{fig:muKdi} at $\Kt = 4.2~K$, we reproduce the $T=0$
behavior of \Fig{fig:phasediagram5.5}.
From this perspective it becomes clear
that the ``CFL island'', which we have found there  
between the {\tscbec} and the {\tscbcs} phase,
corresponds to the upper end of the {\cflbec} regime
and will disappear at a slightly higher value of $\Kt$. 

\begin{figure*}
\subfigure[\;$\Kt = 0$]{
  \includegraphics[width=0.31\textwidth]{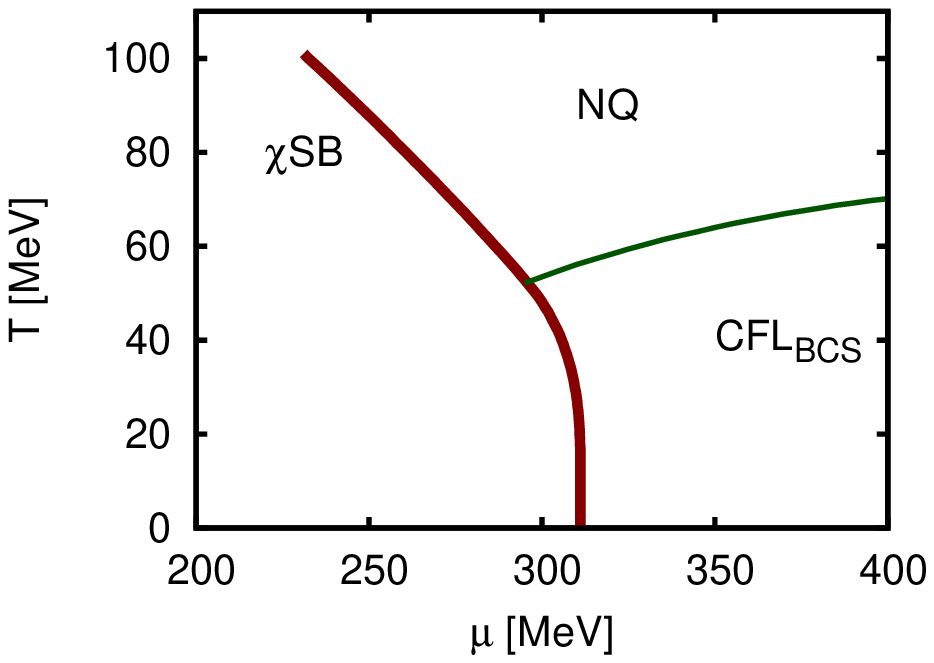}
  \label{fig:Kdi0_5}
}
\subfigure[\;$\Kt = 3.0\,K$]{
  \includegraphics[width=0.31\textwidth]{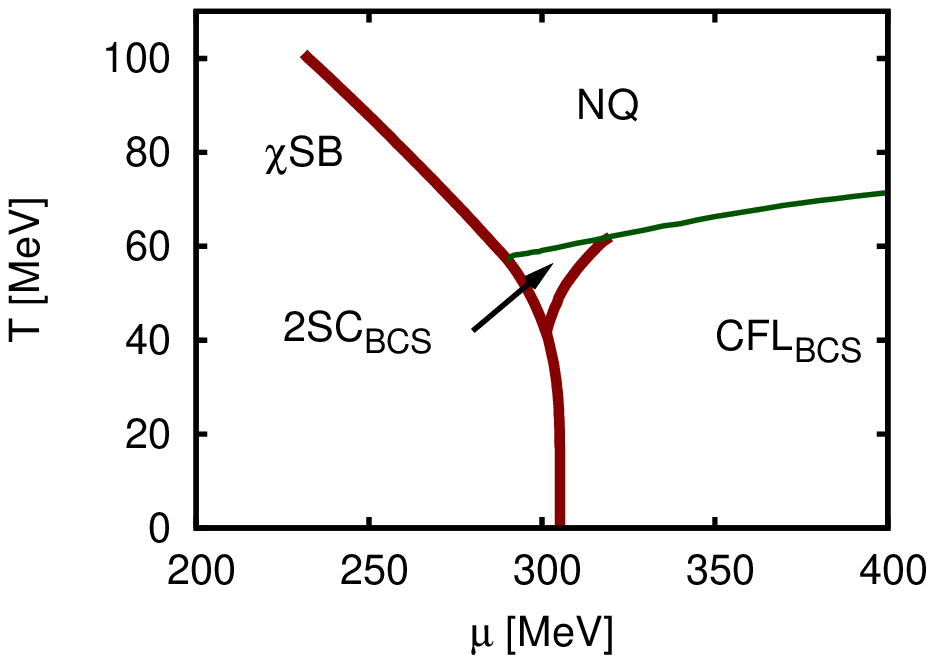}
  \label{fig:Kdi30_5}
}
\subfigure[\;$\Kt = 3.5\,K$]{
  \includegraphics[width=0.31\textwidth]{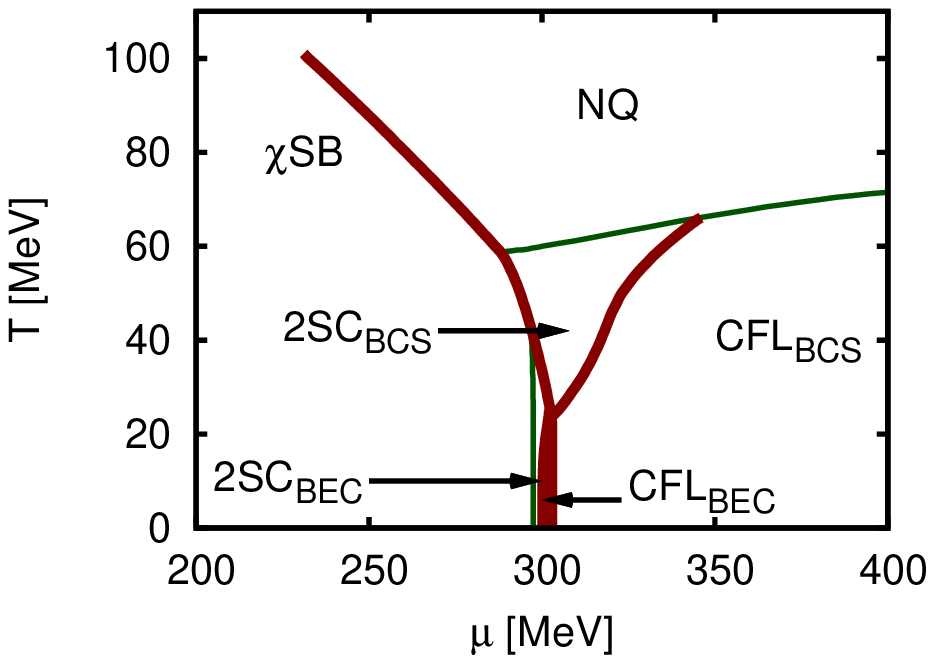}
  \label{fig:Kdi35_5}
}
\subfigure[\;$\Kt = 3.744\,K$]{
  \includegraphics[width=0.31\textwidth]{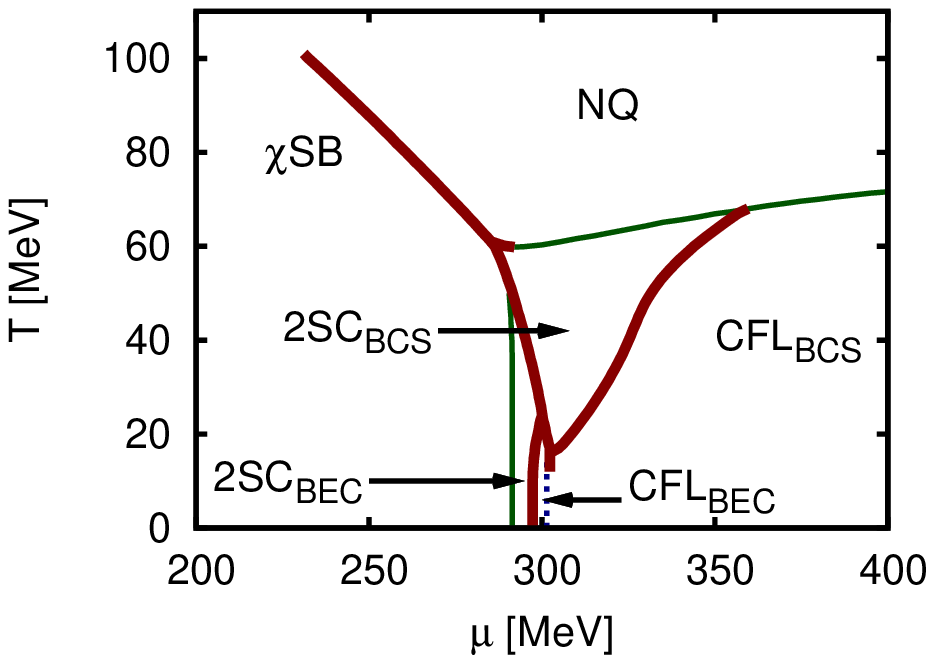}
  \label{fig:Kdi3744_5}
}
\subfigure[\;$\Kt = 4.2\,K$]{
  \includegraphics[width=0.31\textwidth]{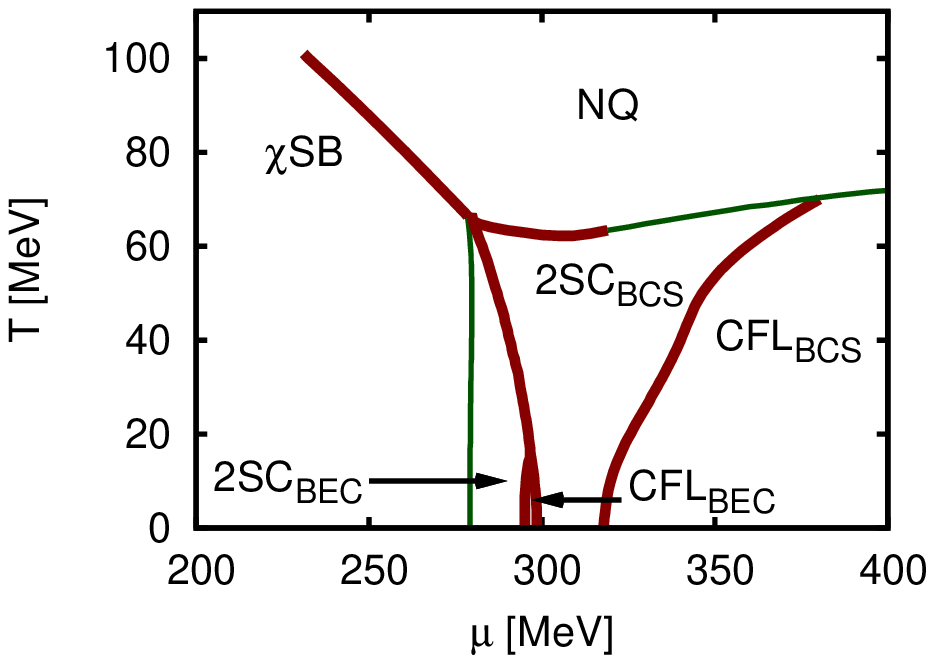}
  \label{fig:Kdi42_5}
}
\subfigure[\;$\Kt = 4.5\,K$]{
  \includegraphics[width=0.31\textwidth]{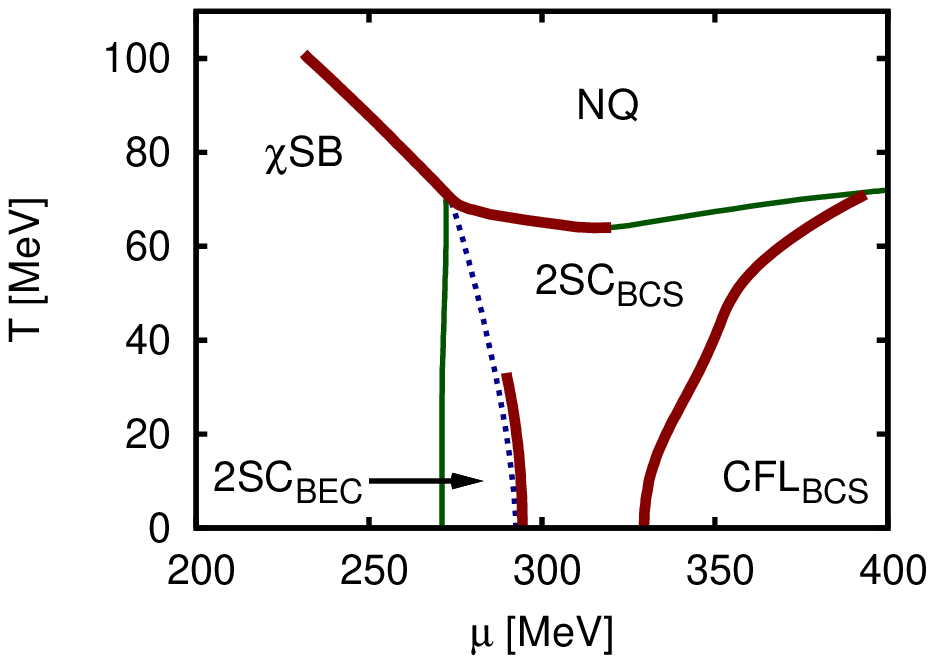}
  \label{fig:Kdi45_5}
}
\caption{The $\mu$-$T$-phase diagram for different choices of $\Kt$
  and equal bare quark masses of ($m_u = m_d = m_s = 5.5\MeV$). 
  The meaning of the different line types is the same as in 
  \Fig{fig:phasediagramCFL}. In the 2SC phase
  the BEC-BCS crossover line (blue dotted line) is defined
  by the condition $M_{u,d}(T,\mu) = \mu$.}
\label{fig:phasediagrams5}
\end{figure*}

In \Fig{fig:phasediagrams5} we show a series of phase diagrams in the 
$\mu$-$T$ plane in order to illustrate how the phase structure evolves 
with $\Kt$. 
While the effect is almost negligible for small and moderate values of 
$\Kt$, a major restructuring takes place at $\Kt \gtrsim 3\,K$:
It starts with a {\tscbcs} phase, which appears near the triple point of 
the three original phases (b) and then grows towards lower temperatures 
and higher chemical potentials. Slightly below $\Kt = 3.5\,K$,
Bose condensation of diquarks sets in, leading to a {\tscbec} phase and
shortly afterwards to a {\cflbec} phase (c). 
At the beginning, the latter is completely separated from the {\cflbcs}
phase by a first-order phase boundary, 
which turns into a crossover upon further increasing $\Kt$.
As a result we find at $\Kt = 3.744\,K$ a critical endpoint (d), which is 
of the same origin as the low-temperature critical endpoint in 
\Fig{fig:phasediagramCFL}.
Note, however, that the endpoint is now located in a region which is
separated from the low-density regime by a first-order phase transition
between 2SC and CFL phase. 
Moreover, the endpoint exists only in an extremely small interval of 
the coupling $\Kt$: While for $\Kt = 3.743\,K$ the 
{\cflbec}-{\cflbcs} phase transition is still first order down to $T=0$, 
already at $\Kt = 3.745\,K$ the entire phase boundary including
the endpoint is covered by the 2SC phase.  
Eventually, the latter reaches the $\mu$-axis and the two CFL regimes 
become separated again (e).
At even higher $\Kt$, the first-order {\tscbec}-{\tscbcs}
phase transition turns into a crossover, leading to another endpoint 
(f).\footnote{
As mentioned earlier, the BEC-BCS cross-over line (blue dotted line),
which we have defined as the line where the non-strange constituent
quark masses are equal to $\mu$, does not exactly run into the 
endpoint. In this sense, the first-order phase transition (related
to a discontinuity in the condensates) does not exactly coincide
with the BEC-BCS transition, although both are closely related to 
each other.}
This endpoint is located at the upper temperature end 
of the first-order boundary, like the ``conventional'' endpoint of the 
$\chi$SB-NQ phase boundary, and it moves downwards with increasing 
$\Kt$. Finally, slightly above $\Kt = 5\,K$, it reaches the $\mu$-axis
and only a crossover remains (not shown). 
On the other hand, as argued before, the 2SC-CFL phase transition is 
necessarily first order, and therefore there is always a discontinuous 
phase transition at low temperatures.

After getting this overview, we would like to understand how
the 2SC phase can be preferred over the CFL phase in some regions of
the phase diagram although we have chosen equal bare quark masses. 
It clearly means that the $SU(3)$-flavor symmetry is spontaneously 
broken, so there is in fact a continuous set of degenerate ground states,
which are related to each other by flavor rotations. 
Bearing in mind that we will introduce larger strange quark masses
later on, we choose, without loss of generality, the standard 2SC 
pairing pattern, i.e., $s_3\not=0$, $s_1=s_2 =0$.

As obvious from Figs.~\ref{fig:muKdi} and \ref{fig:phasediagrams5},
the emergence of the 2SC phase at equal bare quark masses is a
consequence of the six-point interaction $\Lschid$. 
This term couples the chiral condensates to the diquark
condensates and, thus, the constituent quark masses to the diquark 
gaps, see \Eqs{eq:mass} and (\ref{eq:delta}).
However, whereas in the CFL phase this happens symmetrically for all flavors,
in the 2SC phase these equations conspire in a rather peculiar way: 
Since $s_3\not=0$ but $s_1=s_2 =0$, there is a contribution
from the diquark condensates to $M_s$ but not to $M_u$ and $M_d$.
Hence, even if we start with equal bare quark masses, 
in the 2SC phase the strange quarks will be heavier than the non-strange 
quarks. In turn, the larger value of $M_s$ leads to an enhancement of
$\Delta_3$ via an increased modulus of the
(negative) condensate $\phi_3$ in \Eq{eq:delta}.

\begin{figure}
  \includegraphics[width=0.5\textwidth]{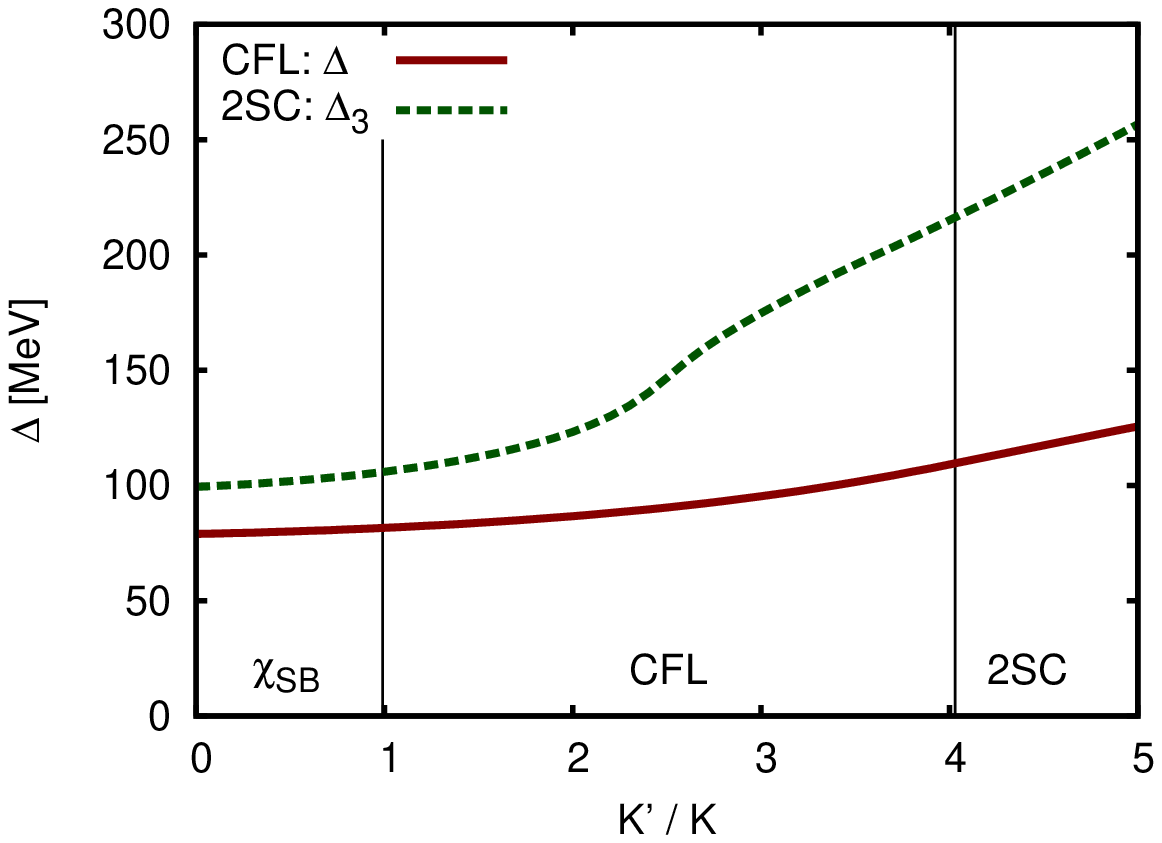}
  \includegraphics[width=0.5\textwidth]{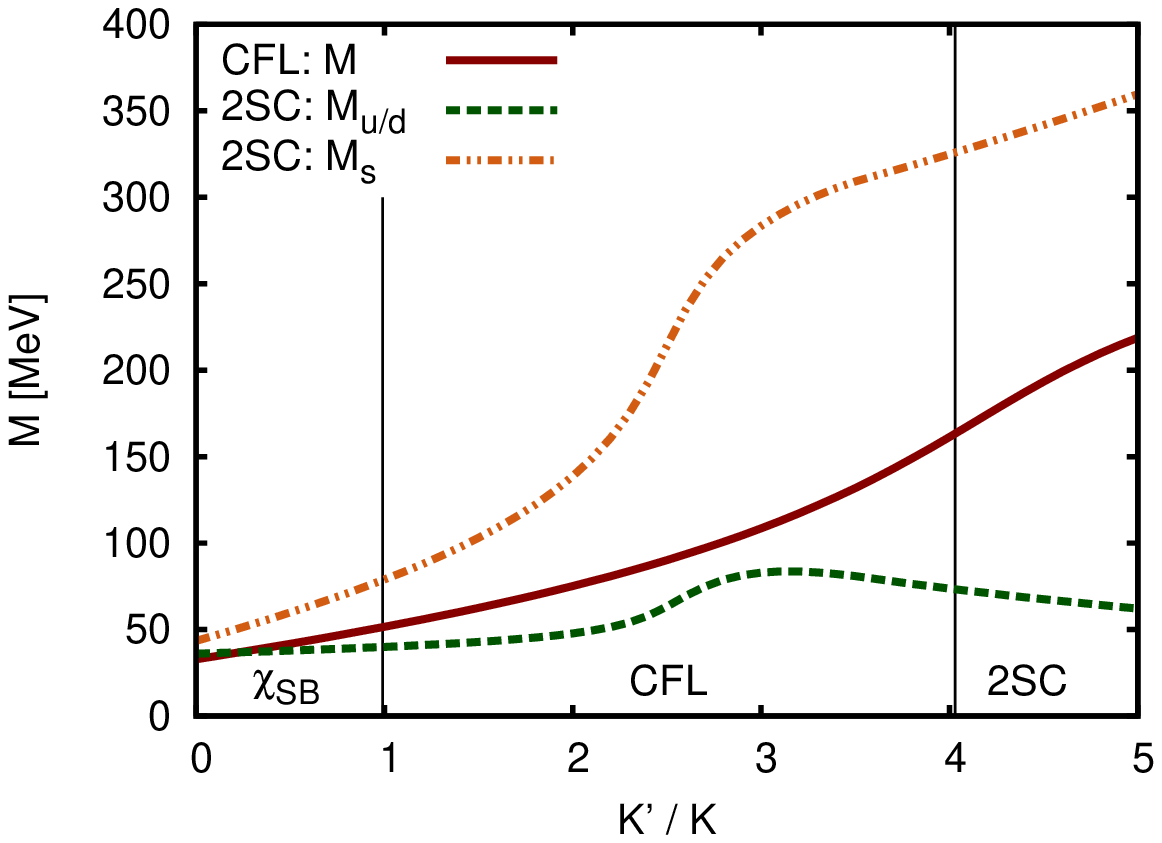}
  \caption{The gap parameters $\Delta_i$ (upper panel) and the 
    constituent quark masses $M_i$ (lower panel)
    in the 2SC and CFL solutions 
    as functions of $\Kt$ for $m_u = m_d = m_s = 5.5\MeV$
    at $T=0$ and $\mu = 310\MeV$.
    The vertical lines indicate the values of $\Kt$ where the
    phase transitions from $\chi$SB to CFL or from CFL to 2SC take
    place (cf.~\Fig{fig:muKdi}).
}
  \label{fig:increasingKdi}
\end{figure}

This effect is illustrated in \Fig{fig:increasingKdi}, where
the diquark gap parameters and the constituent quark masses
are shown as functions of the six-point coupling strength $\Kt$
for the CFL and 2SC solutions at $T=0$ and $\mu=310\MeV$.
One can clearly see that $\Delta_3$ and $M_s$ in the 2SC phase 
rise much faster
with $\Kt$ than the other quantities shown in the figure. 
In particular the ratio between $\Delta_3$ in the 2SC phase and the 
common gap parameter $\Delta$ in the CFL phase rises from about
1.25 at $\Kt=0$ to 2.0 at $\Kt=4$.
It is therefore plausible that eventually the gain in free energy is
larger for 2SC pairing than for CFL pairing. 
In fact, if we had only diquark condensates and no dynamical quark
masses, the pairing energy would be proportional to $\Delta_i^2$ for 
each quasiparticle mode with pairing gap $\Delta_i$. 2SC pairing
would then be favored for 
$\Delta_3|_\mathrm{2SC} > \sqrt{3} \Delta|_\mathrm{CFL}$.
Although in our case the dynamical quark masses (and their
coupling to the diquark condensates) are crucial and certainly 
cannot be neglected,
this estimate can at least roughly explain the numerical results.

At first sight the results found in this section seem to contradict the
GL anlysis of Refs.~\cite{Hatsuda:2006ps,Yamamoto:2007ah, Baym:2008me}
where no 2SC-like solution was found for three equal flavors. 
However, although starting from a general GL potential,
the authors have chosen a particular ansatz for the condensates 
which does not allow for such solutions.
By making a more general ansatz one can show that our results are indeed
consistent with the GL analysis. This is discussed in more detail
in App.~\ref{app:GL}.

\subsection{Realistic strange quark mass}
\label{ssec:strangequarkmass}
\begin{figure}
  \includegraphics[width=0.5\textwidth]{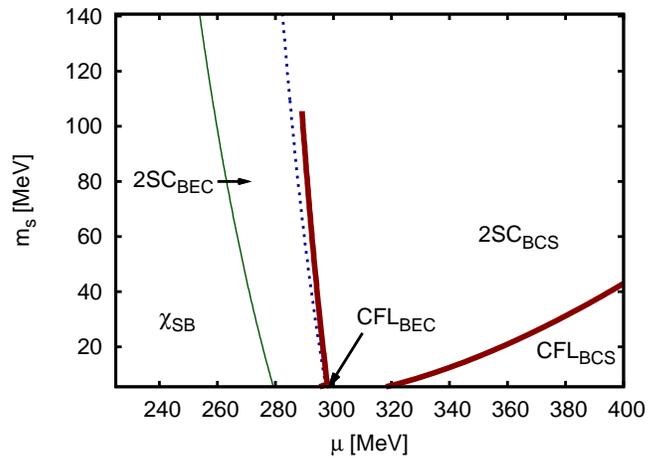}
  \caption{The phase diagram in the $\mu$-$m_s$ plane at $T=0$ for $\Kt
    = 4.2\,K$. The coupling $G$ is adjusted with changing $m_s$
    as described in Sec.~\ref{ssec:parameters}.  
    Thick (red) lines denote first order phase transitions, thin
    (green) lines second order phase transitions.
    The blue dotted line indicates the BEC-BCS crossover line defined
    by $M_{u,d}(T,\mu) = \mu$.}
  \label{fig:mums}
\end{figure}

We now introduce bare strange quark masses that are larger than
the masses of the up and down quarks. 
In particular we are interested in $m_s = 140.7\MeV$, which 
is the value obtained in \Ref{Rehberg:1995kh} by fitting vacuum meson 
properties.
We begin, however, with a more systematic investigation of the 
mass effects by gradually increasing $m_s$ from the equal-mass case
$m_s = 5.5\MeV$ to the ``realistic'' value $m_s = 140.7\MeV$.
Thereby we adjust the coupling $G$ for each value of $m_s$ as
described in Sec.~\ref{ssec:parameters}.

The phase diagram in the $\mu$-$m_s$ plane at $T=0$ 
for $\Kt = 4.2~K$ is shown in \Fig{fig:mums}.
As usual, we now define the CFL phase as a phase where all three
diquark gaps $\Delta_i$ take nonvanishing values but do not need
to be equal.
With rising $m_s$, since pairs involving strange quarks become 
increasingly disfavored, the CFL phase gets more and more replaced by
the 2SC phase, thus pushing the phase boundary to higher values 
of $\mu$.
As the spontaneous breaking of flavor $SU(3)$ through the mechanism 
described above is now stabilized by an explicit symmetry breaking,
even a small enhancement of $m_s$ has a rather large effect. 
For similar reasons, the small CFL area 
between the {\tscbec} phase and the {\tscbcs} phase disappears already
at $m_s \approx 7.5\MeV$.  
Above this point there is a first-order phase transition between the
two 2SC phases, which ends at $m_s \approx 105\MeV$.

A larger value of $m_s$ also leads to a larger value of the 
strange chiral condensate $\phi_3$,
which in turn leads to an enhancement of the effective four-point
quark-quark coupling in the $ud$ channel.   
The latter comes about from closing a strange-quark loop in the
six-point interaction term $\Lschid$ 
and adding this contribution to the genuine four-point vertex from
$\Lfd$. 
As a consequence, the diquarks get bound more deeply, i.e., 
the diquark mass is reduced and, hence, their condensation, which
determines the transition from the $\chi$SB phase to the {\tscbec} phase 
is shifted to lower quark chemical potential.

\begin{figure}
  \includegraphics[width=0.5\textwidth]{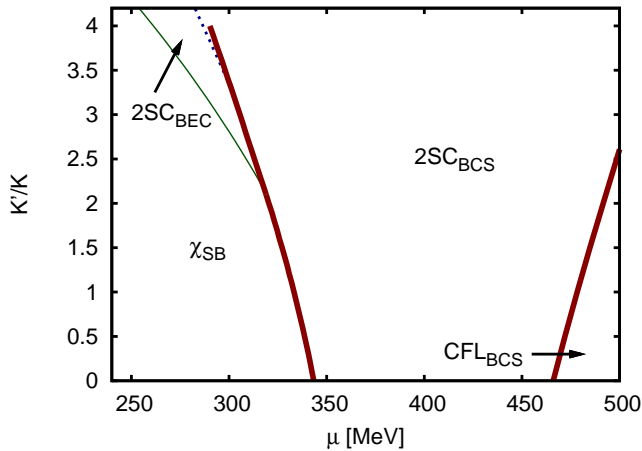}
  \caption{The phase diagram in the $\mu$-$\Kt$ plane at $T=0$ for
    $m_s = 140.7\MeV$. 
    The meaning of the different line types is the same as in 
    \Fig{fig:mums}.}
  \label{fig:Ktmu}
\end{figure}

In the remaining part of this article, we fix the strange quark mass
at the ``realistic'' value, $m_s = 140.7\MeV$. 
In \Fig{fig:Ktmu} we show the phase structure at T = 0 in dependence 
of the six-point coupling strength $\Kt$. With increasing $\Kt$ 
the effective quark-quark coupling becomes stronger and the 
$\chi$SB-2SC phase transition moves to lower quark chemical
potentials. Around $\Kt = 2.2\,K$ the
{\tscbec}  phase forms at the low chemical potential side of the 
chiral phase transition line. 
With larger $\Kt$ this region becomes broader and at $\Kt =
4.0\,K$ the first-order
phase transition between the {\tscbec} phase and the 
{\tscbcs} phase ends.
Similar to the equal-mass case we also find that an increase of 
$\Kt$ is more favorable for the 2SC phase than for the CFL phase,
so that the latter gets shifted to higher values of $\mu$.

\begin{figure*}
\subfigure[\,$\Kt = 0$]{
  \includegraphics[width=0.31\textwidth]{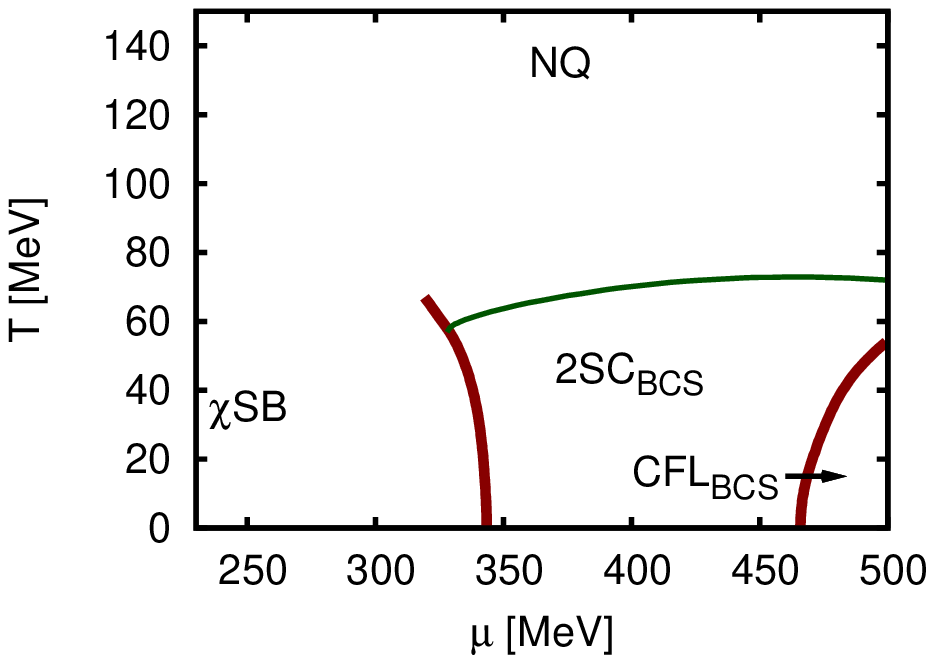}
  \label{fig:Kdi000}
}
\subfigure[\,$\Kt = 1.0\,K$]{
  \includegraphics[width=0.31\textwidth]{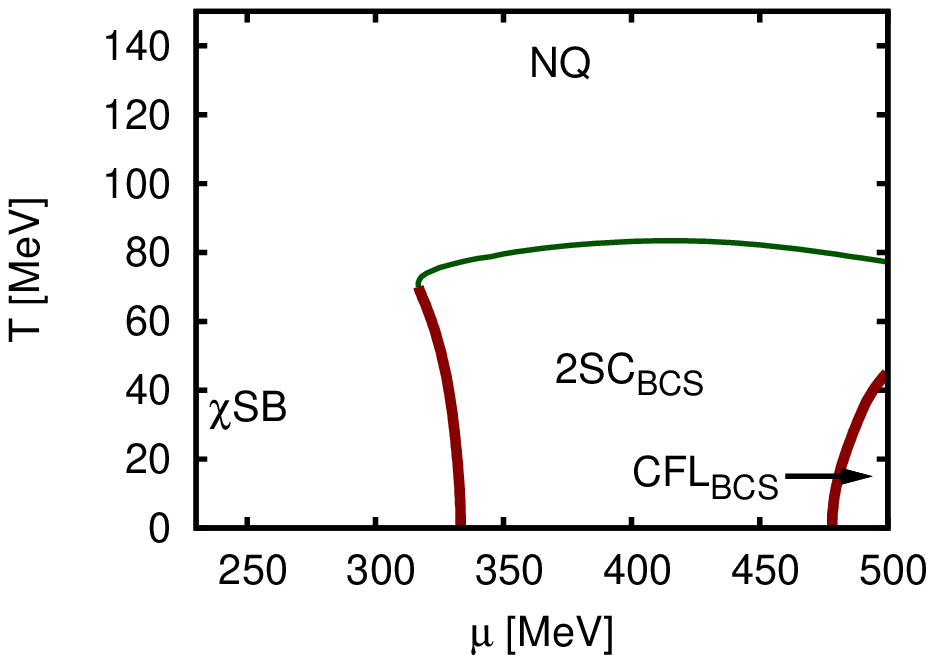}
  \label{fig:Kdi100}
}
\subfigure[\,$\Kt = 2.5\,K$]{
  \includegraphics[width=0.31\textwidth]{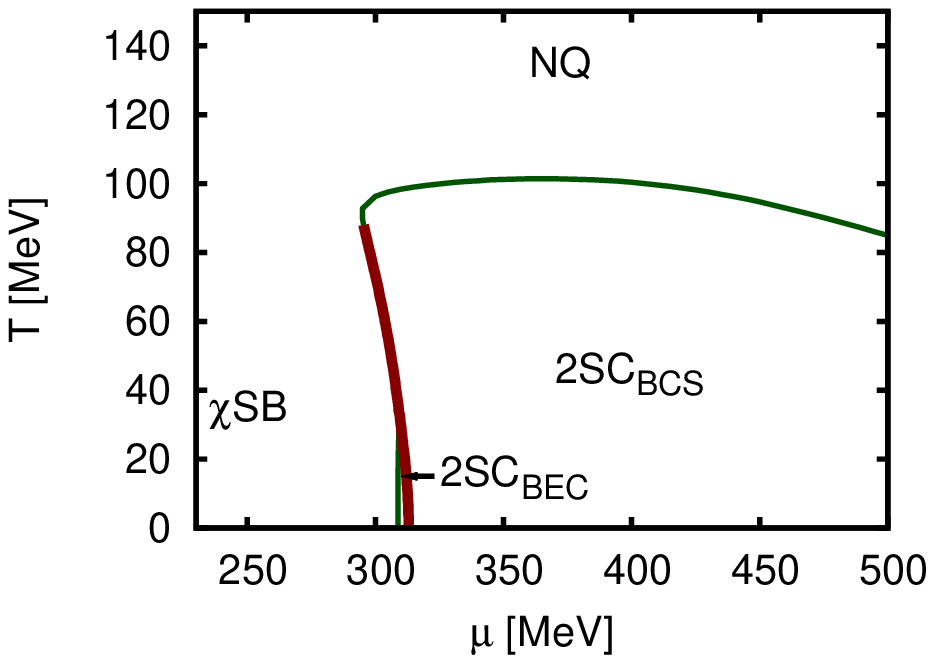}
  \label{fig:Kdi250}
}
\subfigure[\,$\Kt = 3.0\,K$]{
  \includegraphics[width=0.31\textwidth]{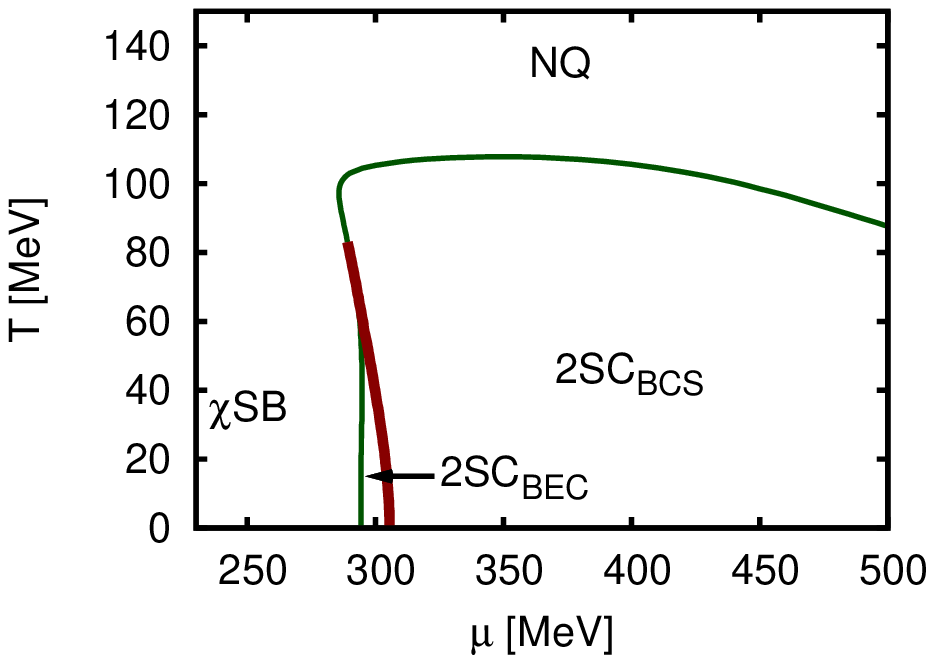}
  \label{fig:Kdi300}
}
\subfigure[\,$\Kt = 3.5\,K$]{
  \includegraphics[width=0.31\textwidth]{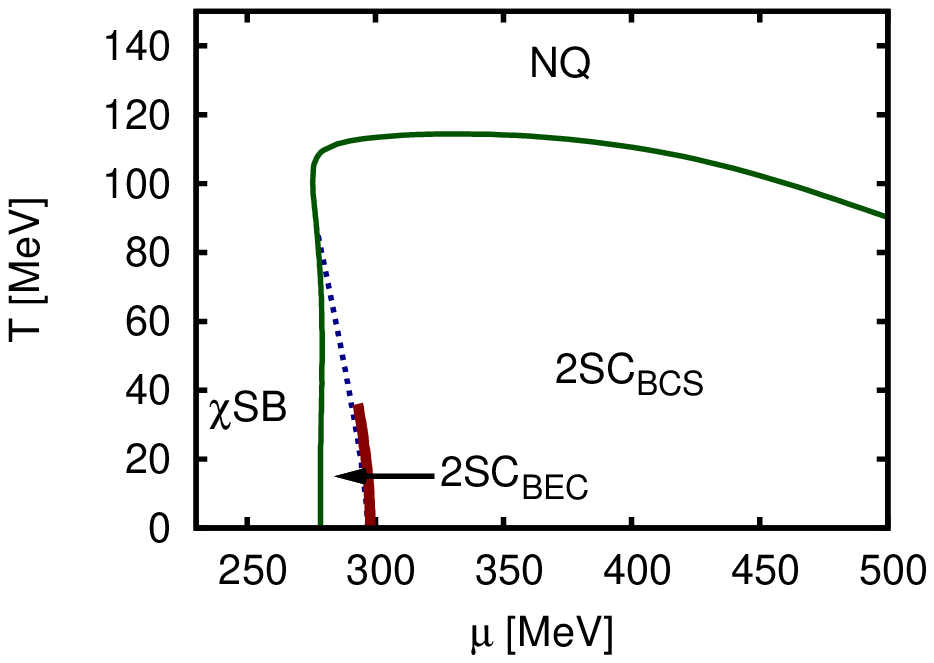}
  \label{fig:Kdi350}
}
\subfigure[\,$\Kt = 4.2\,K$]{
  \includegraphics[width=0.31\textwidth]{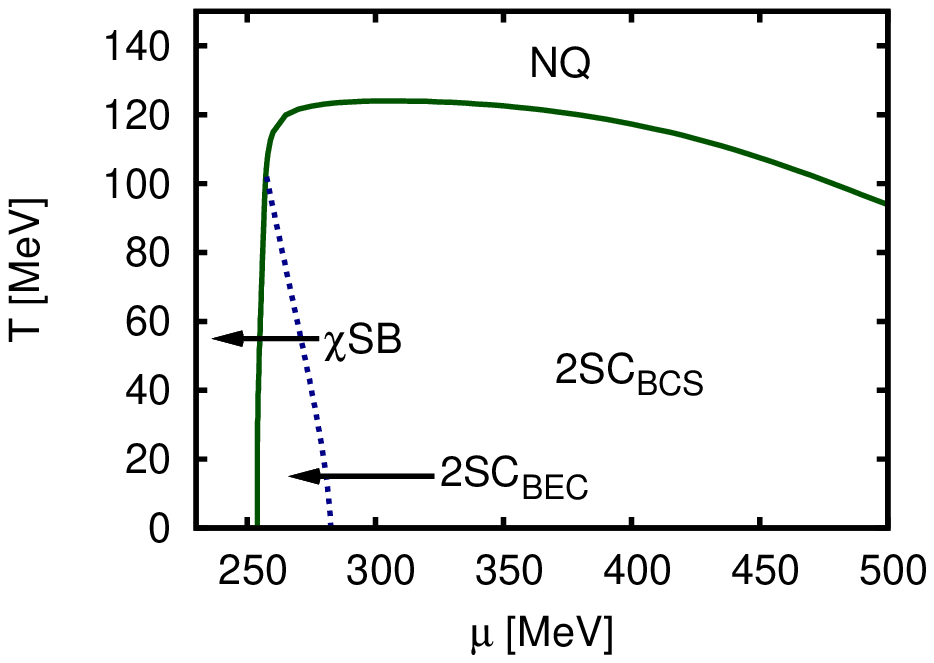}
  \label{fig:Kdi420}
}
\caption{The ($\mu-T$) - phase diagram for different choices of $\Kt$
  and a bare strange quark mass of $m_s = 140.7\MeV$. 
  The meaning of the different line types is the same as in \Fig{fig:mums}.}
\label{fig:phasediagramms}
\end{figure*}

In \Fig{fig:phasediagramms}, we finally show
a series of phase diagrams in the $\mu$-$T$ plane for 
different values of $\Kt$. 
A general result is that none of them contains a critical endpoint
at the low-temperature side of the first-order chiral phase transition.
For $\Kt = 0$ we recover the standard phase diagram (a), which has been
calculated, e.g., in Ref.~\cite{Oertel:2002pj} for the same parameters:
At low temperature chiral symmetry is restored in a first-order 
phase transition between $\chi$SB and 2SC phase, which is continued
by a first-order phase transition between $\chi$SB and NQ phase
at higher temperatures and finally ends in a critical endpoint. 
Furthermore, there is a CFL phase at low temperature and high chemical 
potential. 
With increasing $\Kt$ the 2SC pairing becomes strengthened so that the
CFL phase is pushed to higher chemical potentials while the
first-order chiral phase transition gets successively ``swallowed'' 
by the expanding 2SC phase: As a first step the upwards-moving
second-order 2SC-NQ phase boundary reaches the critical endpoint of the 
$\chi$SB-NQ phase boundary (b).
Next, a {\tscbec} phase emerges on the left-hand-side of the 
$\chi$SB-2SC phase boundary (c,d) and 
eventually the condensation line unites with the
{\tscbcs}-NQ phase boundary (e). 
At the same time first-order chiral symmetry restoration line becomes
disconnected and ends inside the 2SC phase.
When $\Kt$ is further increased, the endpoint moves downwards in 
temperature and finally disappears completely (f).

\section{Conclusions}
\label{sec:conclusions}

We have investigated the phase structure of strongly interacting
matter within a three-flavor NJL-type model with an extended six-point 
interaction that couples chiral and diquark condensates.
This interaction term, which is usually neglected in NJL-model studies
of the phase diagram, has been introduced in \Ref{Abuki:2010jq}
and mimics effects of the axial anomaly.
While this in principle leads to a more complete picture, 
the corresponding coupling constant $\Kt$ is basically unknown
and was therefore treated as a free parameter in our analysis.

Aim of the present study was to extend the investigations of 
\Ref{Abuki:2010jq} to realistic strange quark masses.
In this context we have generalized the mean-field ansatz to
allow for flavor-dependent chiral and diquark condensates. 
This opens the possibility for 2SC pairing, which was
expected to become relevant at large strange quark masses. 

It turned out, however, that even for equal bare quark masses 
the 2SC phase is present in the phase diagram if the
coupling $\Kt$ is sufficiently strong.
This spontaneous breaking of the $SU(3)$ flavor symmetry 
occurs because the axial anomaly induces
a mutual amplification of the strange chiral condensate 
and the non-strange diquark condensate.  
As an important consequence,
we have not found a continuous transition from the low-density
chirally broken phase to the CFL phase at $T=0$ 
for any value of $\Kt$ between $\Kt = 0$ and $\Kt = 5\,K$.
Related to this,
the low-temperature critical end point which was found in
\Ref{Abuki:2010jq} 
only survives in an extremely narrow $\Kt$ interval
and is otherwise covered by the 2SC phase.

As expected, the 2SC phase plays an even more important role at
larger bare strange quark masses, in particular for the
``realistic value'', taken from a fit to vacuum observables.
By varying $\Kt$ we find several qualitatively different phase
diagrams, 
where the first-order chiral phase transition ends outside, inside
or on the phase boundary of the 2SC phase or where 
there is no first-order chiral phase transition at all.
It should be noted, however, that most of these qualitative changes
only occur at relatively large values of $\Kt$, which may turn out 
to be unrealistic. 
But even for smaller couplings, the anomaly can be quantitatively
important, as it stabilizes the 2SC pairing and shifts the CFL phase
to higher chemical potentials.

The present study may be seen as a minimal extension of the analysis
of \Ref{Abuki:2010jq} to include realistic strange quark masses. 
There are, however, many other aspects which have not yet been taken 
into account. Perhaps most important is the consideration of
inhomogeneous phases. For a two-flavor NJL model without 
color superconductivity it has been shown that the entire first-order
chiral phase transition line is covered by an inhomogeneous region
and therefore removed from the phase 
diagram~\cite{Nickel:2009ke,Nickel:2009wj,Carignano:2010ac}.
It would be interesting to see how this result is modified when
strange quarks and color superconducting phases are included and
how the results depend on the axial anomaly.

We have also neglected the possibility of kaon condensation in the CFL
phase~\cite{Schafer:2000ew},
which was found to be very important in an NJL model with realistic 
strange quark masses~\cite{Warringa:2006dk,Basler:2009vk}.
In these references, on the other hand, the anomaly effects described by
$\Lschid$ have not been taken into account.
This term should lead to a higher kaon mass and therefore suppress 
its condensation.
It would be interesting to study this in more detail. 

Finally, we should recall that we have restricted ourselves to
a single quark chemical potential. Similar investigations of 
anomaly effects should also be performed for electrically and color
neutral matter.

\subsection{Acknowledgments}
The work of H.B. was supported by the Helmholtz International Center
for FAIR and by the Helmholtz Graduate School for Hadron and Ion
Research. M.B. acknowledges partial support by EMMI.

\appendix
\section{Ginzburg-Landau approach}
\label{app:GL}

The most general expression for the GL free energy up to order four 
is given in the equations (7), (13) and (14) in \Ref{Yamamoto:2007ah}
as the difference to the normal phase
\begin{equation}
  \label{eq:GL_Omega}
  \Omega\left(\Phi,d_L,d_R\right) = \Omega_\chi\left(\Phi\right) +
  \Omega_d\left(d_R,d_L\right) + \Omega_{\chi d}\left(\Phi,d_R,d_L\right)
\end{equation}
with
\begin{align}
\label{eq:GL_Omega_chi_d_chid}
  \Omega_\chi =& \frac{a_0}{2} \Tr[\Phi^\dagger\Phi] + \frac{b_1}{4!}
  \left(\Tr[\Phi^\dagger
    \Phi]\right)^2+\frac{b_2}{4!}\Tr[(\Phi^\dagger \Phi)^2] 
  \notag\\
  &-\frac{c_0}{2}\left(\det[\Phi] + \det[\Phi^\dagger]\right)\,, 
  \notag\\[1mm]
  \Omega_d =&\alpha_0 \Tr[d_L d_L^\dagger + d_R d_R^\dagger] 
  \notag\\ 
  &+ \beta_1 \left((\Tr[d_L d_L^\dagger])^2 + (\Tr[d_R d_R^\dagger])^2 \right) 
  \notag\\ 
  &+\beta_2 \Tr[(d_L d_L^\dagger)^2 + (d_R d_R^\dagger)^2] 
  \notag\\
  &+\beta_3 \Tr[(d_R d_L^\dagger)(d_L d_R^\dagger)] + \beta_4 \Tr[d_L
  d_L^\dagger]\,\Tr[d_R d_R^\dagger]\, 
  \notag,\\[1mm]
  \Omega_{\chi d} 
  =& \gamma_1 \Tr[(d_R d_L^\dagger) \Phi + (d_L d_R^\dagger)\Phi^\dagger]
  \notag,\\ 
  &+ \lambda_1 \Tr[(d_L d_L^\dagger) \Phi \Phi^\dagger 
   + (d_R d_R^\dagger )\Phi^\dagger \Phi] 
  \notag,\\ 
  &+ \lambda_2 \Tr[d_L d_L^\dagger + d_R d_R^\dagger] \cdot 
     \Tr[\Phi^\dagger \Phi]
  \notag\\
  &+\lambda_3\left(\det[\Phi] \cdot \Tr[(d_L d_R^\dagger) \Phi^{-1}] 
   + h.c.\right)\,.
\end{align}
Here $\Phi$, $d_L$ and $d_R$ are $3\times3$ matrices 
containing the chiral and the left- and right-handed diquark fields,
respectively (for details, see \Ref{Yamamoto:2007ah}).
This ansatz has 13 unkown coefficients which are in general $T$ and
$\mu$ dependent functions and cannot be determined within the GL analysis. 

For the investigation of three massless flavors the authors of
\Ref{Yamamoto:2007ah} have therefore restricted themselves to an
ansatz with equal condensates for all flavors, $\Phi =
\text{diag}\left(\sigma,\sigma,\sigma\right)$ and $d_L = -d_R =
\text{diag}\left(d,d,d\right)$. 
The free energy then takes the simplified form
\begin{align}
\label{eq:GL_3F}
\Omega_{3F}\left(\sigma,d\right) &= \left(\frac{a}{2} \sigma^2 -
  \frac{c}{3} \sigma^3 + \frac{b}{4} \sigma^4\right) \notag\\
&+\left(\frac{\alpha}{2} d^2 + \frac{\beta}{4} d^4\right) 
- \gamma d^2\sigma + \lambda d^2 \sigma^2
\end{align}
with only seven independent coefficients,\footnote{In addition, a term
of order $\sigma^6$ must be introduced by hand in order to stabilize
the system if $b<0$. For simplicity, we neglect such terms in the
present qualitative discussion.}
which are related to the original GL coefficients in 
\Eq{eq:GL_Omega_chi_d_chid} by
\begin{align}
\label{eq:coeff_3F}
a &= 3 a_0\,, &&c = 3 c_0\,,\qquad
b = \frac{1}{2}\left(3b_1 + b_2\right)\,,\notag\\
\alpha &= 12 \alpha_0\,,&&\beta = 12\left(6\beta_1 + 2\beta_2 +
  \beta_3 + 3\beta_4\right)\,,\notag\\
\gamma&=6\gamma_1\,,
&&\lambda = 6\left(\lambda_1 + 3\lambda_2 - \lambda_3\right)\,.
\end{align}
Yet, the full exploration of the remaining seven-dimensional
parameter space would still be almost impossible.
However, by making additional, physically motivated, assumptions
(e.g., positivity of $c$, $\beta$, $\gamma$, and $\lambda$) 
interesting non-trivial results have been extracted in
Refs.~\cite{Hatsuda:2006ps, Yamamoto:2007ah, Baym:2008me}.

Similarly, two-flavor systems have been investigated in~\Ref{Yamamoto:2007ah}
choosing
$\Phi = \text{diag}\left(\sigma,\sigma, 0\right)$ and 
$d_L = -d_R = \text{diag}\left(0,0,d\right)$,
i.e., by completely neglecting all condensates which involve strange quarks.
The resulting free energy is given by
\begin{align}
\label{eq:GL_2F}
\Omega_{2F} \left(\sigma,d\right) 
&= \left(\frac{a^\prime}{2}\sigma^2 +
  \frac{b^\prime}{4} \sigma^4\right) +
\left(\frac{\alpha^\prime}{2} d^2 + \frac{\beta^\prime}{4}d^4 \right) 
+ \lambda^\prime d^2 \sigma^2,
\end{align}
with five independent coefficients,
which are different from those in \Eq{eq:GL_3F}.
In this context it is important to note that by neglecting the strange
condensates completely, it was assumed that the strange quarks are
infinitely heavy and therefore decouple from the non-strange sector.
In particular, a direct comparison of $\Omega_{2F}$ with $\Omega_{3F}$
in order to study the competion between 2SC and CFL pairing was not 
intended and is not possible. 

In contrast, in our NJL-model anlysis, we have studied a 2SC phase 
in a three-flavor environment with a large strange quark chiral condensate.
The corresponding GL ansatz is 
$\Phi = \text{diag}\left(\sigma, \sigma, \sigma_s\right)$ and 
$d_L = -d_R = \text{diag}\left(0,0,d\right)$,
yielding 
\begin{align}
\label{eq:GL_2SC}
&\Omega_{2SC} \left(\sigma,\sigma_s,d\right) =
\notag \\
&\left(
\frac{a}{2}\frac{2\sigma^2 + \sigma_s^2}{3} 
-\frac{c}{3} \sigma^2 \sigma_s 
+ \frac{b_1^{\prime}}{4}\sigma^4 
+ \frac{b_2^{\prime}}{4}\sigma_s^4 
+ \frac{b_1^{\prime} - 2 b_2^{\prime}}{2} \sigma^2 \sigma_s^2
 \right)
\notag\\
& + \left(\frac{\alpha}{6} d^2 + \frac{\beta^\prime}{4} d^4\right)
-\frac{\gamma}{3} d^2 \sigma_s 
+\lambda_1^{\prime} d^2 \sigma^2 
+\lambda_2^{\prime} d^2 \sigma_s^2\,,
\end{align}
where the coefficients $a$, $c$, $\alpha$, and $\gamma$ 
are the same as in \Eq{eq:coeff_3F}, and
\begin{align}
\label{eq:coeff_2SC}
b_1^{\prime} &= \frac{1}{3}(2b_1 + b_2)\,, &&
b_2^{\prime} = \frac{1}{6}(b_1 + b_2)\,,
\notag\\
\beta^\prime &= 4\left(2\beta_1 + 2 \beta_2
+ \beta_3 + \beta_4\right)\,,&&
\notag\\
\lambda_1^{\prime} &= 2\left(2\lambda_2 - \lambda_3\right)\,,&&
\lambda_2^{\prime} = 2\left(\lambda_1 + \lambda_2\right)\,.
\end{align}
Hence, instead of five GL coefficients as in \Eq{eq:GL_2F},
$\Omega_{2SC}$ depends on nine independent coefficients.
In particular, unlike $\Omega_{2F}$, it contains the anomaly induced 
$c$- and  $\gamma$-terms which are related to the six-point
interactions \Eqs{eq:L6} and (\ref{eq:L6tilde}) in the NJL-Lagrangian.

Obviously, the extra terms in \Eq{eq:GL_2SC} are due to the strange
chiral condensate $\sigma_s$, which is not present in $\Omega_{2F}$. 
We can thus recover \Eq{eq:GL_2F} by setting $\sigma_s$ equal to zero.
In fact, a more appropriate way to describe the decoupling
of the strange quarks with large masses is to treat $\sigma_s$ as a
nonvanishing constant. The $\gamma$-term, among others, then leads to
a renormalization of the $d^2$-coefficient $\alpha$, while, e.g., the 
$b_1'$-term leads to a shift of the vacuum energy. This underlines again
that a diract comparison of the free energies obtained with $\Omega_{2F}$
and $\Omega_{3F}$ would be meaningless.

On the other hand, keeping $\sigma_s$ as a dynamical variable, a comparison
of $\Omega_{3F}$ with $\Omega_{2SC}$ is possible.\footnote{Here we tacitly
assume that the GL anaysis is meaningful in both phases, i.e., that all
condensates are small. Clearly, this does not need to be the case.}
To that end, we should replace the coefficient $b$
in \Eq{eq:GL_3F} by the combination $3(b_1'-b_2')$
and the coefficient $\lambda$ by $3(\lambda_1' + \lambda_2')$. 
Moreover, we have to take into account that the coefficients $\beta$ and 
$\beta'$ are linear independent (see \Eqs{eq:coeff_3F} and  
(\ref{eq:coeff_2SC})). This can be accounted for by writing
$\beta = 3(3\beta_1'+ \beta_2')$ and $\beta' = \beta_1'+ \beta_2'$
with $\beta_1' = 4(2\beta_1+\beta_4)$ and $\beta_2' = 4(2\beta_2+\beta_3)$. 
Thus, in oder to compare the free energies of the 2SC and CFL solutions,
we have to deal with ten independent parameters\footnote{A 
straightforward way to obtain a unified description of both, CFL and 
2SC phase, is to make the ansatz 
$\Phi = \text{diag}\left(\sigma, \sigma, \sigma_s\right)$ and 
$d_L = -d_R = \text{diag}\left(d',d',d\right)$. In this case there 
are eleven independent coefficients. Our approach is more
restrictive and therefore has one parameter less.}
($a$, $c$, $b_1'$, $b_2'$, $\alpha$, $\beta_1'$, $\beta_2'$, $\gamma$,
$\lambda_1'$, $\lambda_2'$), plus possible six-order terms which might be
needed for stability reasons.
An exhaustive GL analysis is therefore practically impossible. 

\begin{figure}
  \includegraphics[width=0.5\textwidth]{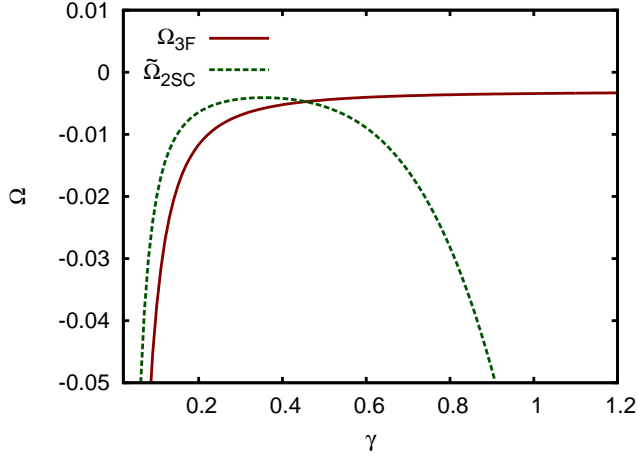}
  \caption{The Ginzburg-Landau free energies of CFL solutions (solid)
           and 2SC solutions (dashed) as functions of $\gamma$ at the 
           endpoint found in 
           \Ref{Hatsuda:2006ps, Yamamoto:2007ah, Baym:2008me} for the 
           example described in the text.}
  \label{fig:GLexampleOmega}
\end{figure}
\begin{figure}
  \includegraphics[width=0.5\textwidth]{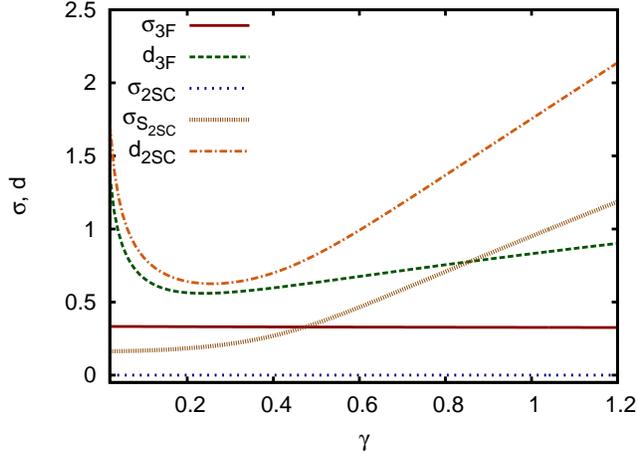}
  \caption{The condensates which minimize the GL free energies in
    the examples given in~\Fig{fig:GLexampleOmega}}
  \label{fig:GLexampleCondensates}
\end{figure}

However, since we only want to show that 2SC-like solutions
can indeed be favored in some cases, it is sufficient to demonstrate
this by an example. In order to proceed, we therefore make a few
simplifying assumptions:
(i) We neglect all $\lambda$-terms, i.e, $\lambda_1=\lambda_2=\lambda=0$.
This was also done in \Ref{Yamamoto:2007ah} in the three-flavor case.
(ii) In the NJL model without the anomaly terms and without diquark coupling,
i.e., taking into account only the interaction term $\Lfchi$, the different
flavors do not mix. Hence their contributions to the thermodynamic potential
are additive. To reproduce this feature we choose $b_1' = 2b/3$, $b_2'=b/3$.
(iii) In the diquark sector we choose $\beta^\prime = \beta / (3\cdot2^{2/3})$. 
In the absence of chiral condensates (or for $\gamma=0$)
this leads to the relation $d_{2SC} = 2^{1/3}\,d_{3F}$, 
in agreement with weak-coupling QCD~\cite{Rischke:2003mt}. 
As can be seen in \Fig{fig:increasingKdi}, this relation is also 
approximately fulfilled in the NJL model at $\Kt=0$. 
(iv) We restrict the possible 2SC solutions to BCS-like solutions, i.e., 
we set $\sigma=0$ in \Eq{eq:GL_2SC}. 

For the 2SC phase, we then obtain the simplified GL potential
\begin{align}
\label{eq:GL_2SCtilde}
&\tilde\Omega_{2SC} \left(\sigma_s,d\right) =
\notag \\
&\left(
\frac{a}{6}\sigma_s^2  
+ \frac{b}{12}\sigma_s^4 \right)
+ \left(\frac{\alpha}{6} d^2 + \frac{\beta}{12\cdot2^{2/3}} d^4\right)
-\frac{\gamma}{3} d^2 \sigma_s \,,
\end{align}
while the CFL phase is described by \Eq{eq:GL_3F} with $\lambda=0$. 
Thereby both potentials depend on the same coefficients
so that the resulting free energies can be compared with each other. 
In order to further reduce the number of parameters, we perform this
comparison at the location of the low-temperature critical end point found
in \Ref{Hatsuda:2006ps, Yamamoto:2007ah, Baym:2008me},
\begin{equation}
\label{eq:aalpha}
a=\frac{c^2}{3b} + \frac{2\gamma^2}{\beta}\,,\qquad
\alpha = -\frac{\beta c^2}{27\gamma b^2}\,.
\end{equation}
This eliminates $a$ and $\alpha$ from the equations,
so that we are left with four parameters, $b$, $c$, $\beta$, and $\gamma$.

Finally, we introduce an arbitrary scaling factor $\Lambda$ of 
dimension energy and measure all dimensionful quantities in units of the 
corresponding power of $\Lambda$. We then simply choose 
$b = c = \beta = 1$ in these units and study the different free-energy
solutions at the point given by \Eq{eq:aalpha} for varying values of 
$\gamma$. This anaysis has been done numerically.

In \Fig{fig:GLexampleOmega} we show the results for $\tilde\Omega_{2SC}$ 
and $\Omega_{3F}$ for our example. We see that the CFL solution is favored
at low values of $\gamma$ whereas the 2SC solution becomes favored at 
larger values of $\gamma$.
This phase transition is in qualitative agreement with our findings 
in the NJL model when $\Kt$ is increased, cf. \Fig{fig:muKdi}.

The values of the condensates associated with in minima of the
free energy are shown in \Fig{fig:GLexampleCondensates}. 
Their behavior with rising $\gamma$ is qualitatively similar to
the effect of an increased $\Kt$ in the NJL calculations,
cf.~\Fig{fig:increasingKdi}. In both cases when increasing the
coupling ($\gamma$ or $\Kt$) the diquark and the strange quark chiral
condensate in the 2SC phase are rising much faster than the
condensates in the CFL phase.

The NJL-model results are thus completely consistent with a GL analysis
if the latter is performed with a sufficiently general ansatz for the
condensates.


\end{document}